\RequirePackage{fix-cm}
\documentclass[smallcondensed]{svjour3}     % onecolumn (ditto)
\smartqed  % flush right qed marks, e.g. at end of proof
\usepackage{graphicx}
\usepackage{epstopdf}
\usepackage{epsfig}
 \pdfoutput=1

\RequirePackage{fix-cm}

\usepackage[numbers]{natbib}
\usepackage{bm}
\usepackage{amsmath,amssymb}
\usepackage{epstopdf}
\usepackage{epsfig}
\usepackage{booktabs}
\usepackage{rotating}
\usepackage{makecell}
\usepackage{multirow}
\usepackage{color}
\usepackage{enumitem}
\newlist{steps}{enumerate}{1}
\setlist[steps, 1]{label = Step \arabic*:}
\setlist[enumerate]{itemsep=0mm}

\def\square{{\vcenter{\vbox{\hrule height.3pt
        \hbox{\vrule width.3pt height5pt \kern5pt
           \vrule width.3pt}
        \hrule height.3pt}}}}

\def\tlint{{- \kern-0.85em \int \kern-0.2em}}
\def\dlint{{- \kern-1.05em \int \kern-0.4em}}

\begin{document}

\title{
Threshold-Based Portfolio: The Role of the Threshold and Its Applications
%Threshold-based portfolio: tailoring portfolios to specific risk-return profiles using thresholds for %forecasts of stock returns
%Insert your title here%\thanks{Grants or other notes
}


\titlerunning{Threshold-Based Portfolios}        % if too long for running head

\author{Sang Il Lee         \and
        Seong Joon Yoo %etc.
}


\institute{Sang Il Lee \at
              Department of Computer Engineering, Sejong University,
	 Seoul, 05006, Republic of Korea \\
              \email{silee@sejong.ac.kr}           %  \\
           \and
           Seong Joon Yoo \at
              Department of Computer Engineering, Sejong University,
	 Seoul, 05006, Republic of Korea \\
	          %Tel.: +82-2-3408-4468\\
              %Fax: +82-2-3408-4321\\
              \email{sjyoo@sejong.ac.kr}
}

\date{Received: date / Accepted: date}

\maketitle

\begin{abstract}
This paper aims at developing a new method by which to build 
a data-driven portfolio featuring a target risk-return.
%We define the investment universe of 10 stocks in $S\&P500$ and 
%select a long short-term memory (LSTM) model as the best resulting model for one-month-ahead return forecasts %through a comparative study on the accuracy of recurrent neural network models (RNNs).  
%accuracy for future
%using recurrent neural networks (RNNs): a simple RNN, long short-term memory (LSTM), and %gated recurrent units (GRU).
%a fully data-driven method
%for asset allocations based on stock return prediction.
%Recurrent Neural Networks (RNNs) are applied to stock return predictions. 
%We build portfolios by adjusting potential return threshold levels used to classify assets, and obtain risk-return profiles examining their statistical properties.
We first present a comparative study of recurrent neural network models (RNNs), including
a simple RNN, long short-term memory (LSTM), and gated recurrent unit (GRU) for selecting
the best predictor to use in portfolio construction.
The models are applied to the investment universe consisted of ten stocks in the $S\&P500$.
The experimental results shows that LSTM outperforms the others in terms of hit ratio of one-month-ahead forecasts.
We then build predictive threshold-based portfolios (TBPs) that are
subsets of the universe satisfying given threshold criteria for the predicted returns.
The TBPs are rebalanced monthly to restore equal weights to each security within the
TBPs.
We find that the risk and return profile of the realized TBP 
represents a monotonically increasing frontier on the risk--return plane, 
where the equally weighted portfolio (EWP) of all ten stocks plays a role in
their lower bound. This shows the availability of TBPs in 
targeting specific risk--return levels, and 
an EWP based on all the assets plays a role in the reference portfolio of TBPs.
In the process, thresholds play dominant roles in characterizing risk, return, 
and the prediction accuracy of the subset.
The TBP is more data-driven in designing portfolio target risk and return 
than existing ones, in the sense that it requires  
no prior knowledge of finance such as financial assumptions, 
financial mathematics, or expert insights.
In a practical application, we present the TBP management procedure 
for a time horizon extending over multiple time periods; we also discuss 
their application to mean--variance portfolios to reduce estimation risk.


%Portfolio, Investment, Recurrent neural networks, Financial time series, %Prediction accuracy, Heuristic efficient frontier 
    
\end{abstract}

\section{Introduction}
\label{intro}
%Your text comes here. Separate text sections with
Today, machine learning has come to play an integral role 
in many parts of the financial ecosystem, from portfolio management and 
algorithmic trading, to fraud detection and 
loan/insurance underwriting.  
Time series are one of the most common data types 
encountered in finance, and so time-series analysis is 
one of the most widely used traditional approaches in finance and economics.
The development of machine learning algorithms has opened
a new vista for modeling the complexity of financial time series
as an alternative to the traditional econometric models, by effectively combining
diverse data and capturing nonlinear behavior. 
For this reason, financial time-series modeling 
has been one of the most interesting topics that has arisen in the
application of machine learning to finance.
Researchers have successfully modeled financial time series 
by focusing primarily on prediction accuracy or automatic trading rules
\cite{ats09,dix15,huc09,huc10,kra17,mor14,ser13,tak13,cav16,agg17,gao16,zha15,fis17, pan18, zin18,sin17}. 
Nevertheless, 
financial modeling and applications remain daunting, given
the difficulties arising from the 
essentially nonlinear, complex, and evolutionary 
characteristics of the financial market.

On the other hand, asset allocation has been traditionally 
considered an issue central to investment and risk management.
Markowitz \cite{mar52} was the first to introduce 
a rigorous mathematical framework for allocation, 
called modern portfolio theory (MPT). 
Based on a mean--variance optimization technique,
MPT provides a method by which to assemble 
assets and maximize the expected return of the portfolio for a given level of risk.
Following Markowitz's thinking, new portfolio models
have been subsequently proposed for more practical use
and to achieve a better understanding of portfolios. Examples include
the thee-factor asset pricing model \cite{fam92}, 
the Black--Litterman model\cite{bla92},
the resampled efficient frontier model \cite{mic98},
the global minimum variance model \cite{hau91},
the maximum diversification portfolio \cite{cho08},
and the risk-parity portfolio \cite{qia05,qia09}.
Additionally, dynamic/tactical asset allocations based on simple rules or 
market anomalies
were developed to automatically adjust portfolios in response to market changes 
\cite{fab14,kel14,kel14_2,kel16}.
These studies show that today there is general consensus about 
the importance of effective combinations of
assets.
 
In this study, we propose a new method for constructing a data-driven portfolio 
using recurrent neural networks (RNNs)-based future return predictions.
Throughout this study, 
we will refer to this portfolio as a threshold-based portfolio (TBP), 
since its properties are characterized 
by the threshold levels imposed on predicted returns.
In particular, this study makes the following main contributions to the literature:
\begin{itemize}[label=$\bullet$]
\item It examines the ability of RNNs to forecast one-month-ahead stock returns.
\item It develops a new TBP portfolio method and analyzes their properties.
\begin{itemize}[label=$\bullet$]
\item 
The threshold can be used for a parameter to draw a TBP frontier that 
comprises the set of TBP points on a risk--return plane. 
This implies that one can build a portfolio with a specific risk--return 
level by selecting the appropriate threshold level.
\item The equally weighted portfolio (EWP) is the lower bound of the TBPs on the TBP frontier.
This implies that the TBPs can be characterized with respect to a reference portfolio, EW.   
\end{itemize}
\item In practical application, it develops the management process for TBPs 
for pursuing specific risk--return levels over multiple-periods 
and for the method incorporating TBPs into MPT.
\end{itemize}

The remainder of this paper is organized as follows:
Section \ref{sec_relatedwork} discusses
some of the important work related to this area. 
Section \ref{prediction models} explains the simple 
recurrent neural network (S-RNN), long short-term memory (LSTM), 
and gated recurrent units (GRU).
Section \ref{results} provides experimental results regarding 
the prediction accuracy 
of the models and the performance of TBPs. In Section \ref{sec_app},
we discuss the practical applications of TBPs. Finally, in Section \ref{conclusion},
we conclude this paper and discuss possible future extensions of our work.

\section{Related Work}\label{sec_relatedwork}
We present LSTM-based predictions and prediction-based portfolios.
\subsection{Financial Time Series Prediction Using RNNs}
Using conventional econometric models,
financial economists have found there to be 
statistically significant relationship 
between stock returns and lagged variables.
For example, 
Campbell et al. \cite{cam93} investigated the relationship between aggregate stock market
trading volume and the serial correlation of daily stock returns.
They provide an evidence that a stock price
decline on a high-volume day is more likely than a stock price decline on a
low-volume day to be associated with an increase in the expected stock return. 
Choueifaty and Coignard \cite{cho00} show that trading volume 
is a significant determinant of the lead-lag patterns observed in stock returns. 

For this reason, we select RNN algorithms; these
are superior for modeling time-lag effects 
in multi-dimensional financial time series, by virtue of feeding 
the network activations from a previous time step as inputs into
the network, to influence predictions in the current step.
In contrast, feed forward neural networks (FFNNs) are not appropriate 
for capturing these time-dependent dynamics: they 
operate on a fixed-size time windows, and so
they can provide only limited temporal modeling. 

RNNs are less commonly applied to financial time-series 
predictions, yet some recent studies 
has shown promising results for use in financial time-series prediction.
Fischer and Krauss \cite{fis17} 
%Long short-term memory (LSTM) networks are a 
%state-of-the-art technique for sequence learning.
%They are less commonly applied to fnancial time series 
%predictions, yet inherently suitable for this
%domain. 
deployed LSTM networks to predict 
one-day-ahead directional movements in 
a stock universe and constructed subset portfolios 
by selecting constituents outperforming the
cross-sectional median return of all stocks in the next day. 
They found that LSTM networks outperform memory-free
classification methods (i.e., a random forest (RAF), a deep neural net (DNN), and a logistic
regression classifier (LOG)) on measures of the mean return per
day, annualized standard deviation, annualized Sharpe ratio, and accuracy. 

More recently, Bao et al. \cite{bao17} developed a hybrid model
called the WSAEs--LSTM, combined with
wavelet transforms (WT), stacked autoencoders (SAEs), and LSTM 
to effectively combine diverse financial data, including 
historical trading data of open price, high price, low price, closing price, and volume and 
technical indicators of stock market indexes and macroeconomic variables.
The experimental results show that it produces
more accurate one-day-ahead stock price predictions 
than the similar models, including WLSTM 
(i.e., a combination of WT and LSTM), RNN, and LSTM.

\subsection{Machine Learning Prediction-Based Investment Portfolios}
Machine learning  
has been applied to portfolio construction while 
focusing on the portfolio optimization problem,
with multiple objective functions being subject to a set of constraints.
Machine learning-based prediction is a valuable tool 
that can be used to mitigate difficulties inherent in traditional methods,
i.e., ranking stocks and assessing 
their future potential.

Freitas et al. \cite{fre09}
present a new model of 
prediction-based portfolio optimization for capturing short-term investment
opportunities. 
For the universe of Brazilian stocks, they 
combined their neural network predictors featuring normal prediction errors
with the mean--variance portfolio model, and show that
the resulting portfolio outperforms the mean--variance model 
and beats the market index.
More recently, Mishra et al. \cite{mis16} developed a novel prediction 
based mean--variance (PBMV) model, as an 
alternative to the conventional Markowitz mean--variance model, 
to solve the constrained portfolio optimization problem. 
They present a low-complexity heuristic functional link artificial neural network (HFLANN) model
to overcome the incorrect estimation taken as the mean of the past returns 
in the Markowitz mean--variance model, and carry out
the portfolio optimization task by using multi-objective evolutionary algorithms (MOEAs).
Ganeshapillai et al. \cite{gan13} propose a machine learning-based method
to build a connectedness matrix and address
the shortcomings of correlation in capturing
events such as large losses. 
They show that
the matrix can be used to build portfolios
that not only ``beat the market,'' but also
outperform optimal (i.e., minimum-variance)
portfolios.

The results of these studies show that machine learning-based
estimations can be effectively used to overcome certain 
limitations inherent in traditional method.
With regard to prediction-based portfolios, our work
is in line with this thinking, but is more fundamental in the sense that
we focus heavily on predicted returns by imposing thresholds
with respect to prediction 
and diversification effects, by aggregating stocks rather than
adopting existing portfolio frameworks. This fact makes TVP
more data-driven than existing models.

\section{Models: S-RNN, LSTM, GRU}\label{prediction models}

S-RNNs \cite{elm90} are an extension of a conventional FFNN that adds a feedback connection
to a feedback network consisting of three layers: an input layer,
a hidden layer, and an output layer. However, \cite{ben94} found it is difficult 
to train an S-RNN to capture long-term dependencies, because the gradients tend to
either vanish or explode. Alternatively, LSTM \cite{hoc97} and GRU \cite{cho14} 
have been proposed to overcome the problem by using a ``gating'' approach.
The LSTM algorithm   
is local in space and time \cite{hoc97}, which means that the
computational complexity of learning LSTM models per weight and time
step with the stochastic gradient descent (SGD) optimization technique is
O(1), and the learning computational complexity per time step is O(W),
where W is the number of weights. Hence, our model is capable
of handling large-scale data, as the computational complexity of our model
grows linearly with respect to the length of the input data.

\subsection{LSTM Architecture}

LSTMs can effectively learn important pieces of information that 
may be found at different positions in the financial time series,
by controlling what is added and removed 
from memory in the hidden layers. This is conducted by using 
a combination of three gates: (1) a forget gate, (2) an input gate, and (3) an output gate.\\

\noindent {\bf Forget Gate:} An LSTM cell receives the current input $x_{t} \in \mathbb{R}^{d}$, the hidden state vectors $\mathbf{h}_{t-1}\in \mathbb{R}^{n}$, and a cell state $\bm{C}_{t-1} \in \mathbb{R}^{n}$ at time $t-1$. The forget gate then is then calculated as 
\begin{equation}
f_{t}=\sigma(\bm{W}_{f}\mathbf{x}_{t}+\mathbf{U}_{f}\mathbf{h}_{t-1}+\mathbf{b}_{f}),
\end{equation}
where:
\begin{itemize} %[label=$\bullet$]
\item $\mathbf{W}_{f}\in \mathbb{R}^{n\times d}$
is the weight matrix from the input $\mathbf{x}_{t}$ to the forget gate 
$\mathbf{f}_{t}$,
\item $\mathbf{U}_{f}\in \mathbb{R}^{n\times n}$ is the
weight matrix from the previous hidden vector $\mathbf{h}_{t-1}$ to the
forget gate $\mathbf{f}_{t}$, 
\item $\mathbf{b}_{f}\in \mathbb{R}^{n}$ is the forget gate bias, 
\item $\mathbf{f}_{t} \in
\mathbb{R}^{n}$ is the output of the gate, which determines 
the amount to be erased from the previous cell state, and
\item $\sigma(\cdot)$ is a sigma function.\\
\end{itemize}

\noindent {\bf Input Gate:} The input gate $\mathbf{i}_{t}$, which is used to scale the candidate update vector $\widetilde{\mathbf{C}}_{t}\in \mathbb{R}^{n}$, determines what parts of $\widetilde{\mathbf{C}}_{t}$ are added to the corresponding memory cell element at time $t$, based
on the recurrent connection from the hidden vector $\mathbf{h}_{t-1}$ and the input
at time $t$, $\mathbf{x}_{t}$:  
\begin{eqnarray}
\mathbf{i}_{t}&=&\sigma(\mathbf{W}_{i}\mathbf{x}_{t}+\mathbf{U}_{i}\mathbf{h}_{t-1}+\mathbf{b}_{i}),\\
%\end{equation} 
\widetilde{\mathbf{C}}_{t}&=&tanh(\mathbf{W}_{c}\mathbf{x}_{t}+\mathbf{U}_{c}\mathbf{h}_{t-1}+\mathbf{b}_{c}),
\label{eqn_C}
\end{eqnarray}
where:
\begin{itemize}%[label=$\bullet$]
\item $\mathbf{W}_{i}\in \mathbb{R}^{n \times d}$, $\mathbf{U}_{i}\in\mathbb{R}^{n \times n}$, and $\mathbf{b}_{i}\in \mathbb{R}^{n}$ are the input gate parameters,
\item $\mathbf{W}_{c}\in \mathbb{R}^{n \times d}$, $\mathbf{U}_{c}\in \mathbb{R}^{n \times n}$, 
\item $\mathbf{b}_{c} \in \mathbb{R}^{n}$ are the parameters for selecting
a candidate state, $\widetilde{\mathbf{C}}_{t}$, and
\item $tanh(\cdot)$ is the tanh function.
\end{itemize}
Then, the current state of the cell $\mathbf{C}_{t}\in \mathbb{R}^{n}$ is given
by 
\begin{equation}
\mathbf{C}_{t}=\mathbf{i}_{t}\odot\widetilde{\mathbf{C}}_{t}+\mathbf{f}_{t}\odot \mathbf{C}_{t-1},
\end{equation}
where $\odot$ represents the element-wise Hadamard product. \\

\noindent {\bf Output Gate:} The output gate $\mathbf{o}_{t} \in \mathbb{R}^{n}$, which
is used to calculate 
the output 
$\mathbf{h}_{t}\in\mathbb{R}^{n}$, determines the output from the current
cell state:
\begin{eqnarray}
\mathbf{o}_{t}&=&\sigma(\mathbf{W}_{o}\mathbf{x}_{t}+\mathbf{U}_{o}\mathbf{h}_{t-1}+\mathbf{b}_{o}), \\
\mathbf{h}_{t}&=& \mathbf{o}_{t}\odot tanh(\mathbf{C}_{t}),
\label{eqn_h}
\end{eqnarray}
where $\mathbf{W}_{o}\in \mathbb{R}^{n\times d}$, $\mathbf{U}_{o}\in \mathbb{R}^{n\times n}$,
and $\mathbf{b}_{o}\in \mathbb{R}^{n}$ are the output gate parameters.
The hidden vector $\mathbf{h}_{t}$ of the memory cell can be used as the final output of the network. 

\subsection{GRU Architecture}

The structure of a GRU can be expressed as follows: 
\begin{eqnarray}
\mathbf{z}_{t}&=&\sigma(\mathbf{W}_{z} \mathbf{x}_{t}+\mathbf{U}_{z}\mathbf{h}_{t-1}+\mathbf{b}_{z}), \\
\mathbf{r}_{t}&=&\sigma(\mathbf{W}_{r} \mathbf{x}_{t}+\mathbf{U}_{r} \mathbf{h}_{t-1}+\mathbf{b}_{r}), \\
\mathbf{h}_{t}&=&\mathbf{z}_{t}\odot \mathbf{h}_{t-1}+(1-\mathbf{z}_{t})
\odot \textrm{tanh}[\mathbf{W}_{h}\mathbf{x}_{t}+\mathbf{U}_{h}(\mathbf{r}_{t}
\odot \mathbf{h}_{t-1})+\mathbf{b}_{h}],
\end{eqnarray}
where:
\begin{itemize}%[label=$\bullet$]
\item $\mathbf{x}_{t},\mathbf{h}_{t},\mathbf{z}_{t}$, and $\mathbf{r}_{t}$ 
are the input vector, output vector, update gate vector, 
and reset gate vector, respectively, and
\item $\mathbf{W},\mathbf{U}$, and $\mathbf{b}$ are forward matrices, recurrent matrices, and
biases, respectively.
\end{itemize}

\section{Experiment}\label{results}

%In this section, we illustrate the performance of our proposed algorithm RoAdam 
%compared to RLSTM, SR-LSTM, and 
%RN-LSTM on both synthetic data and real time series.

\subsection{Data}\label{data}

\subsubsection{Universe}
\begin{figure}[t]
\centering
  \scalebox{0.4}
  {
	\includegraphics{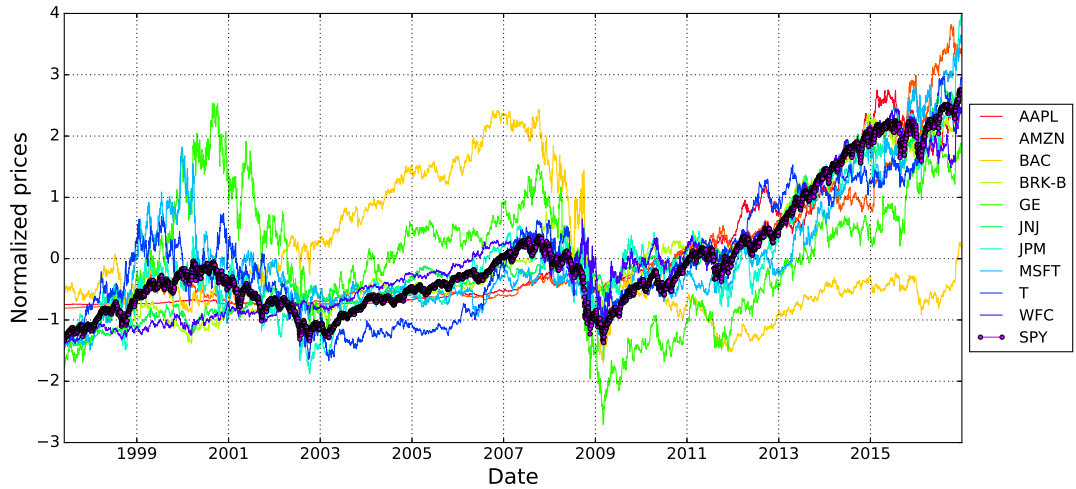}
   }

\caption{(Color online) Normalized stock prices for the 10 sample stocks over the test period
}
\label{fig_stockprice}
\end{figure}

The asset universe consists of the top 10 stocks in Standard and Poor's 500 index (S$\&$P500):
\begin{itemize}[label=$\bullet$]
\item Apple (AAPL), Amazon (AMZN), Bank of America Corporation (BAC), Berkshire Hathaway Inc. Class B (BRK-B), General Electric Company (GE), Johnson$\&$Johnson (JNJ), JPMorgan Chase $\&$ Co. (JPM), Microsoft Corporation (MSFT), AT$\&$T Inc. (T), and Wells Fargo $\&$ Company (WFC).
\end{itemize}
We use data from January $1997$ to December $2016$ from Yahoo Finance.  
The daily stock dataset contains five attributes: open price, high price, 
low price, adjust close price, and volume (OHLCV).
Figure \ref{fig_stockprice} graphically shows the normalized closed price 
(i.e., subtract the mean from each original value and then divide by the standard deviation). 
We convert the daily OHLCV dataset to four different monthly OHLCV datasets by calculating
the last, mean, maximum, and minimum values
of the daily OHLCV dataset per month.
Each monthly OHLCV dataset is used as a raw dataset for forecasting 
one-month-ahead return at the end of each calendar month.

% (to be discussed in the following section).
%%%%%%%%%%%%%%%%%%%%%%%%%%%%%%%%%%%%
\subsubsection{Preprocessing}
 
To achieve higher quality and reliable predictions, the five attributes are preprocesed as a percentage change,  $(x^{(t)}-x^{(t-1)})/x^{(t-1)}$. All data were divided into 
a training dataset (70$\%$) to set the model parameters 
and the test set (30$\%$) for an out-of-sample model evaluation.
The 30$\%$ of the training set is used as the validation to evaluate a given model
during training.

The statistical characteristics of the data used 
to train and test the deep learning model 
are shown in Table \ref{StatData}. 
We observe that the data are roughly in the range of $-1$ to $1$ 
which is a usual range of features in deep learning, save for
abnormal trading volume from max values of the volume data. 

\begin{table}[htbp]
\tiny
  \centering
  \caption{Statistics of data (closed prices and volume) used to train, validate, and test the RNN models}
    \begin{tabular}{lrrrrrrrrrr}
    \toprule
    %      &       &       &       &       &       &       &       &       &       &  \\
    %\midrule
          & AAPL  & AMZN  & BAC   & BRK-B & GE    & JNJ   & JPM   & MSFT  & T     & WFC \\
          \midrule
          & \multicolumn{10}{c}{Train data} \\
    Mean  & 0.01, 0.45 &  0.01  0.12 &  0.01  0.124 &  0.00   0.21 & 0.00   0.17 & 0.007   0.186 &  0.00   0.16 &  -0.00   0.13 & -0.00  0.20 & 0.01  0.13 \\
    Std.   & 0.16,2.22 &  0.20   0.59 & 0.08   0.549 & 0.06   0.79 & 0.07 0.74 & 0.069   0.901 &  0.11   0.83 &  0.12   0.73 &  0.10   0.84 &  0.07 0.62 \\
    Min.   & -0.57, -0.88 & -0.41  -0.75 &  -0.22 -0.76 &  -0.12  -0.76 & -0.17  -0.617 & -0.157 -0.66 &  -0.28  -0.80 &  -0.34  -0.80 &  -0.18  -0.76 & -0.16 -0.84 \\
    Max   & 0.45,16.68 & 0.62   1.71 & 0.17   2.198 & 0.26   2.34 &  0.19   2.93 & 0.174  5.603 & 0.25  5.17 & 0.40   4.84 & 0.29   4.56 & 0.23,  2.93 \\
    \hline
          & \multicolumn{10}{c}{Validation data} \\
    Mean  & 0.06, 0.31 & -0.00   0.11 &  0.01   0.152 & 0.00   0.13 &  0.00   0.082 & 0.00   0.14 &  0.01   0.21 &   0.00   0.32 &  0.015  0.20 & 0.01,  0.12 \\
    Std.   & 0.12, 1.06 & 0.14   0.60 & 0.03  0.822 & 0.02   0.52 & 0.02  0.393 & 0.027   0.68 &  0.04   0.77 & 0.05   1.66 &  0.03   0.878 & 0.02,   0.58 \\
    Min.   & -0.15, -0.72 &  -0.30  -0.70 & -0.04  -0.68 & -0.05  -0.50 &  -0.065  -0.48 & -0.04  -0.65 & -0.06  -0.59 &  -0.11  -0.79 & -0.06  -0.73 &  -0.02,  -0.67 \\
    Max.   &  0.35,   5.03 & 0.36   1.96 &  0.09   3.62 &  0.05   1.30 & 0.058  1.19 & 0.05   2.90 & 0.09   2.85 & 0.08   8.50 &  0.08   4.30 &  0.07,  1.41 \\
    \hline
          & \multicolumn{10}{c}{Test data} \\
    Mean  &  0.03,  0.11 &  0.04   0.36 & -0.00   0.22 &  0.00  0.27 &  -0.011   0.11 & 0.00   0.10 & 0.00   0.20 & 0.00   0.15 & 0.00  0.05 & 0.00,  0.23 \\
    Std.  &  0.12,   0.54 &  0.14  1.40 &  0.21   0.68 & 0.06   1.06 &  0.11   0.51 & 0.04  0.51 & 0.11   0.75 &  0.08   0.70 & 0.06  0.40 & 0.14,   0.80 \\
    Min.   &  -0.32,  -0.82 & -0.25  -0.72 & -0.53  -0.60 &  -0.14  -0.86 & -0.27 -0.58 &  -0.12  -0.61 & -0.23 -0.73 & -0.16  -0.69 & -0.15  -0.63 &  -0.35,  -0.74 \\
    Max.   & 0.23,   1.71 & 0.54   6.45 & 0.73   2.25 &  0.12   5.64 & 0.25  1.87 & 0.07   1.49 & 0.24   2.39 & 0.24   3.05 & 0.09   1.50 & 0.40,  3.65 \\
    \bottomrule
    \end{tabular}%
  \label{StatData}%
\end{table}%

\subsection{Experimental Design}
We build S-RNN, LSTM, and GRU architectures for one-month-ahead forecasts of stock returns. 
Based on the validation set evaluation, we carried out a grid search 
over their hyperparameters over the number of RNN hidden layers (1,2, or 3) and
the number of hidden units per layer (8, 16, 32, 64, or 128), and whether dropout is used 
to avoid overfitting of the model.
The whole networks was trained by a backpropagation algorithm by minimizing the  
the quadratic loss value, $L=\frac{1}{2}\sum_{t}^{T}(r(t)-\hat{r}(t))^{2}$ 
(where $\hat{r}(t)$ is the output of the last layer and $r(t)$ is the corresponding target) 
on the validation set. The efficient ADAM (adaptive moment estimation) 
optimization algorithm \cite{kin14} with a learning
rate of 0.001 is used to fit the models 
in mini-batches of size 20. From the experiments, 
we specified the topology consisting of an input layer, 
an RNN layer with $h=36$ hidden neurons, a 50$\%$ dropout layer, and
an output layer with a linear activation function for regression.

The feature vectors to feed the models are overlapping sequences of 36 consecutive points (trading months in three years) for the preprocessed variables.
The sequences themselves are sliding windows shifted 
by one month for each time $t\geq 36$, that is,
$\{\mathbf{x}_{t-35},\mathbf{x}_{t-34},\cdots,\mathbf{x}_{t} \}$.

The experimental set-up is implemented over a laboratory prototype,
equipped with an Intel quad core i7-6700 processor at 3.4GHz, Nvidia GPU (i.e., GTX 1070), 
and 32GB of RAM running the Ubuntu 16.04.2 LTS x86-64 Linux
distribution. Prediction models are evaluated using Keras 2.0.4 \cite{cho16} and TensorFlow 0.11.0. In our approach, the stage of modeling and forecasting 
stock returns contributes significantly to the overall processing time 
and, for one asset, is obtained in an approximate processing time of
109 seconds.

\subsection{Prediction accuracy}
We evaluate the predictive ability of the three models using the hit ratio, which is defined as follows:
\begin{align}
\textrm{Hit ratio}=\frac{1}{N}\sum_{i=1}^{N} P_{i}(i=1,2,\ldots,N),
\end{align}
where $N$ is total trading months and $P_{i}$ is the prediction result 
for the $i^{th}$ trading day, defined as:
%\begin{equation}
\[ P_{i} =
  \begin{cases}
    1       & \quad \text{if } y_{t+1} \cdot \hat{y}_{t+1}
    >0, \\
    0  & \quad  \text{ otherwise},\\
  \end{cases}
\]
where $y_{t+1}$ and $\hat{y}_{t+1}$ are the realized return at the last 
business day of month $t+1$ 
and the one-month-ahead return predicted at the last 
business day of month $t$, respectively.

% from the ACORN-SAT dataset. 
%We resample the daily OHLCV to four different monthly OHLCV by selecting
%the last, mean, maximum, and minimum values each month, and
%they are used for raw data to forecast one-month-ahead returns at last business day.
%To obtain monthly raw data, we convert the daily stock trading data 
%to monthly data using the last, mean, maximum, and minimum values of each month.
Table \ref{tab_hit}
shows the mean and standard deviation (SD) of the hit ratios 
for the 10 assets and the use of
the last business day and the LSTM model
generates the best prediction accuracy value (0.604).
Therefore, we will use the LSTM model and the last business day OHLCV of each month
for building TBPs in the next subsections.
\begin{table}[htbp]
  \centering
  \caption{Mean and SD of hit ratios for the ten assets, respectively}
    \begin{tabular}{lrrrr}
    \toprule
         &  Last  & Mean  & Max   & Min \\
    \midrule
    S-RNN&  0.559, $\mathbf{0.040}$ & 0.555, 0.066 & 0.555, 0.059  & 0.555, 0.048 \\
    LSTM&  $\mathbf{0.604}$, 0.042 & 0.536, 0.083 & 0.569, 0.048 & 0.584, 0.039  \\  
    GRU&  0.573, 0.053 & 0.550, 0.079 & 0.586, 0.049 & 0.575, 0.051  \\
    \bottomrule
    \end{tabular}%
  \label{tab_hit}%
\end{table}%

\subsection{Role of Threshold and TBP}%Rebalancing rules}

We present the three type of TBP imposing the positive and negative threshold levels 
($\theta^{+}$ and $\theta^{-}$, respectively) on the predicted returns. Given
the one-month-ahead return prediction $\hat{r}_{i}$ for asset 
$S_{i} (i=1,2,\ldots, n)$, the TBPs are defined as 
the subset of the universe:
\begin{itemize}[label=$\bullet$]
\item Long only TBP: $\{ S_{i} \in \textrm{Universe} \textrm{ } |\textrm{ } \hat{r}_{i}\geq \theta^{+} \}$

\item Short only TBP: $\{ S_{i} \in \textrm{Universe} \textrm{ } |\textrm{ } \hat{r}_{i} < -\theta^{-} \}$

\item Long--short TBP: $\{ S_{i} \in \textrm{Universe} \textrm{ } |\textrm{ } \hat{r}_{i}\geq \theta^{+}  \textrm{and }  \hat{r}_{i} < -\theta^{-}  \}$
\end{itemize}
To illustrate, the thresholds are used to classify assets as long and short positions. 
A long (short) equity portfolio consists of assets 
whose prediction is higher (lower) than  $\theta^{+}$ ($\theta^{-}$).   
Here, the thresholds are exogenous variables, and
as explained in the next sections,
we can determine proper threshold values for the target portfolio through backtesting.

\subsection{Portfolio Weight}
As classical portfolio models, the TBPs are built on the following underlying
assumptions for evaluating performance. 
(i) all stocks are infinitely divisible; 
(ii) there are no restrictions on buying and selling any selected portfolio; 
(iii) there is no friction (transactions costs, taxation, commissions, liquidity, etc.); and 
(iv) it is possible to buy and sell stocks at closing prices at any time $t$.

We adapt the periodic rebalancing strategy:
the investor adjusts the weights in his portfolio on the last business day of 
every month, as academic research typically assumes 
monthly portfolio rebalancing.
Throughout this study, we provide 
the results of experiments on the long TBP (TBP in short), and other TVPs can be easily built 
by adjusting the thresholds. Let $w_{i}$ denote the TBP weight on the $i^{\textrm{th}}$ asset.
The TBPs ($w_{i}\geq 0$) are subjected to the budget constraint 
$\sum_{i}^{P}w_{i}=1$, where $P$ is the number of assets in the TBP.
For all the TBPs, $w_{i}$ is defined as $|w_{i}|=1/P$, that is, equally-weighted TBPs.

\subsection{Simulation Results}

\subsubsection{Experiment 1: Performance of TBPs}
Table \ref{Table_performance_TBP} provides the mean and SD 
(standard deviation) values of the monthly returns of the individual assets, 
the EWP of the universe, and the TBPs with different thresholds.
The explanation is as follows:
\begin{itemize}[label=$\bullet$]
\item AAPL and AMZN achieve higher returns with higher volatility.
\item The EWP achieves lower volatility by diversification effect, 
which leads higher Sharpe ratio.
\item The EWP and TBPs overly outperform individual stocks in terms of
the Sharpe ratio.
\item As shown in the TBPs, an increase in $\theta$ results 
in an increase in the return and volatility of TBPs.
\end{itemize}

\begin{table}[htbp]
  \centering
  \caption{Performance of the individual assets, EWP, and TBPs }
    \begin{tabular}{rrrrrrr}
    \toprule
    \multicolumn{2}{c}{Asset/Portfolios} & \multicolumn{1}{c}{Threshold} & Mean  & SD   
     & Mean/SD 
    &  \thead{Average \\ assets}\\
    \midrule
    \multicolumn{2}{l}{AAPL} & \multicolumn{1}{l}{} & 0.02  & 0.075 & 0.271 &\\
    \multicolumn{2}{l}{AMZN} & \multicolumn{1}{l}{} & 0.023 & 0.08  & 0.299 &\\
    \multicolumn{2}{l}{BAC} & \multicolumn{1}{l}{} & 0.013 & 0.102 & 0.128 &\\
    \multicolumn{2}{l}{BRK} & \multicolumn{1}{l}{} & 0.01  & 0.0378 & 0.275 &\\
    \multicolumn{2}{l}{GE} & \multicolumn{1}{l}{} & 0.01  & 0.053 & 0.202 &\\
    \multicolumn{2}{l}{JNJ} & \multicolumn{1}{l}{} & 0.0129 & 0.036 & 0.353 &\\
    \multicolumn{2}{l}{JPN} & \multicolumn{1}{l}{} & 0.014 & 0.075 & 0.19 &\\
    \multicolumn{2}{l}{MSFT} & \multicolumn{1}{l}{} & 0.017 & 0.062 & 0.275 &\\
    \multicolumn{2}{l}{T} & \multicolumn{1}{l}{} & 0.01  & 0.042 & 0.239 &\\
    \multicolumn{2}{l}{WFC} & \multicolumn{1}{l}{} & 0.011 & 0.048 & 0.237 &\\
    \multicolumn{2}{l}{EWP} & \multicolumn{1}{l}{} & 0.014 & 0.039 & 0.368 &\\
    \hline
    \multicolumn{2}{l}{\multirow{5}[2]{*}{TBP}} & \multicolumn{1}{l}{0} & 
    0.015 & 0.04  & 0.381 & 9.072\\
    \multicolumn{2}{l}{} & \multicolumn{1}{l}{0.005} & 0.017 & 0.044 & 0.381 & 6.637  \\
    \multicolumn{2}{l}{} & \multicolumn{1}{l}{0.01} & 0.02  & 0.052 & 0.383 & 4.376\\
    \multicolumn{2}{l}{} & \multicolumn{1}{l}{0.015} & 0.02  & 0.06  & 0.347 & 2.855\\
    \multicolumn{2}{l}{} & \multicolumn{1}{l}{0.02} & 0.018 & 0.069 & 0.269 & 2.173\\
    \bottomrule
    \end{tabular}%
  \label{Table_performance_TBP}%
\end{table}%

The remarkable fact is that the EWP serves as a benchmark
for evaluating the TBPs, in the sense that
there is a (roughly) consistent up--right shift from the point of the EWP
on the risk--return plane: as $\theta$ increases, 
the return increases from 0.014 (EWP)
to 0.015 ($\theta=0.000$), and then to 0.018 ($\theta=0.02)$;
the risk increases from 0.039 (EWP) to 0.04 ($\theta=0$), 
and then to 0.069 ($\theta=0.02$). 
In the next section, this relationship is more clearly elucidated 
as the risk--return frontier.

The EWP has been frequently used as a proxy for the risk--return ratio
of the financial market, by both academia and the financial industry  
 \cite{jeg01,ply01}. 
It is more diversified than a value-weighted portfolio, which is 
heavily concentrated into just the largest companies, so that
it is being widely traded in the real financial industry (e.g.,
the NASDAQ-100 Equal Weighted Index
allots the same weight to each stock in the index). 
Therefore, the fact that such a well-known EWP serves as a benchmark 
for TVPs allows us to more quantitatively characterize TBPs.

\begin{figure}[t]
\centering
  \scalebox{0.3}
  {
	\includegraphics{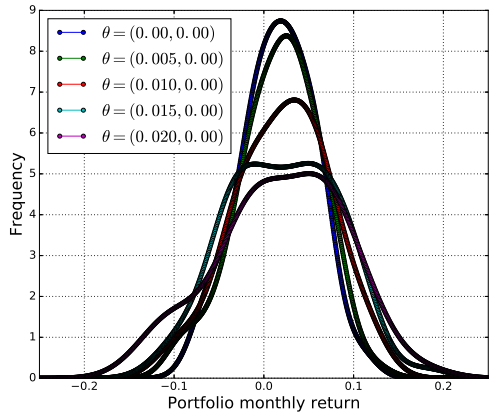}
   }
\caption{(Color online) Distributions of TBP monthly returns}
\label{fig_DP_system}
\end{figure}

\begin{table}[htbp]
  \centering
  \caption{Prediction accuracy values for assets whose predictive return is higher than $\theta$ over the test period}
    \begin{tabular}{lrrr}
    \toprule
    $\theta$ & \thead{No. of correct \\ forecasts} & \thead{No. of total \\ forecasts }  & Accuracy \\
    \midrule
    \multicolumn{1}{l}{0} & 343   & 562   & 0.61 \\
    \multicolumn{1}{l}{0.0025} & 321   & 521   & 0.616 \\
    \multicolumn{1}{l}{0.005} & 225   & 405   & 0.629 \\
    \multicolumn{1}{l}{0.0075} & 204   & 326   & 0.625 \\
    \multicolumn{1}{l}{0.01} & 171   & 272   & 0.628 \\
    \multicolumn{1}{l}{0.0125} & 134   & 216   & 0.62 \\
    \multicolumn{1}{l}{0.015} & 110   & 177   & 0.621 \\
    \multicolumn{1}{l}{0.0175} & 95    & 152   & 0.625 \\
    \multicolumn{1}{l}{0.02} & 83    & 134   & 0.619 \\
    \multicolumn{1}{l}{0.0225} & 74    & 117   & 0.632 \\
    \multicolumn{1}{l}{0.025} & 67    & 108   & 0.62 \\
    \bottomrule
    \end{tabular}%
  \label{Threshold_Accuracy}%
\end{table}%

We examine the relationship between the magnitude of predictive returns and the prediction accuracy. 
Table \ref{Threshold_Accuracy}
shows the correct forecasts among
all forecasts whose value is larger than $\theta$.
For the different $\theta$s, the prediction accuracy ranges over
$0.61 \sim 0.63$, independent of $\theta$.

We calculate the accumulated returns of TVPs  
by using the closing prices on the last trading day of each month. 
We rebalance all portfolios on the last trading day of each month based on the
one-month-ahead prediction; we then reinvest according
to a weight vector that divides the accumulated wealth equally among
the constituents. 
The accumulated return $R_t$ is defined as:
\begin{align}
R_{t} = \prod _{i=0}^{t}(1+r_{i}),
\end{align}
where $r_{i}$
is the arithmetic return at time $i$. This is a standard performance measure
for comparing investments, and it relates the wealth at time $t$, $W_{t}$
, to the initial wealth, $W_{0}$, as 
$W_{t} = W_{0}\times R_{t}$.
All experiments in this study used an initial wealth value of $W_{0} = 1$.
Figure \ref{CumulativeR_Scenario2} shows
the cumulative returns of the individual assets, EWP, and TBPs.
\begin{figure}[t]
\centering
  \scalebox{0.35}
  {
	\includegraphics{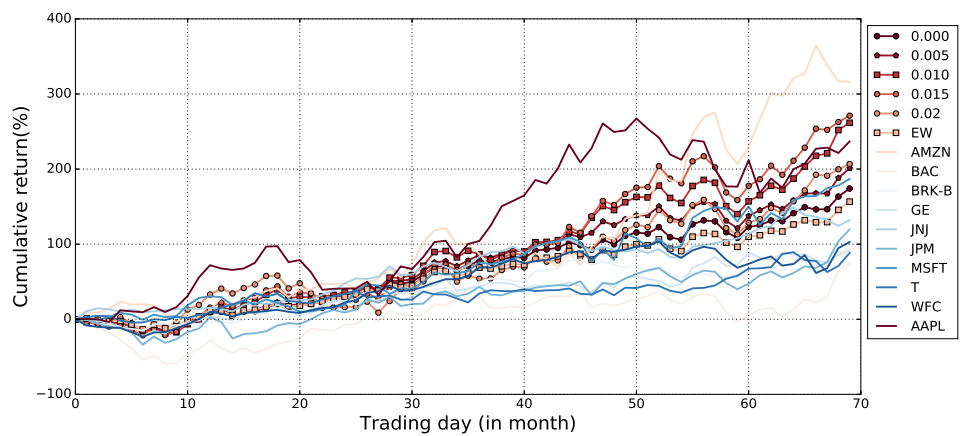}
   }
\caption{(Color online) Cumulative returns of individual assets, EWP, and TBPs}
\label{Fig_cumul_1}
\end{figure}

\subsection{Experiment 2: Robustness Test}
\begin{figure}[t]
\centering
  \scalebox{0.35}
  {
	\includegraphics{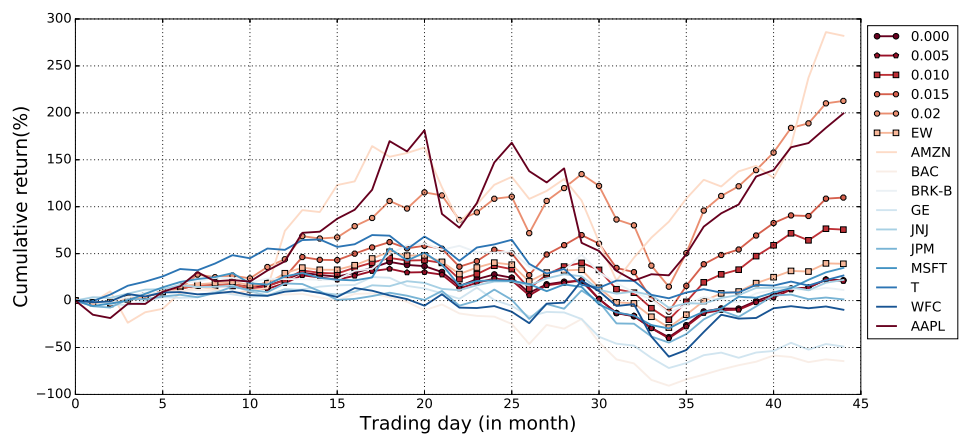}
   }
\caption{(Color online) Experiment 2 results: Cumulative returns of individual assets, EW, and TBPs}
\label{CumulativeR_Scenario2}
\end{figure}

As a further check, we conduct a similar experiment 
with the whole of the study period (i.e., January 1, 2006 to December 31, 2014).
Figure \ref{CumulativeR_Scenario2} graphically shows the cumulative returns over the test period (i.e., 30$\%$ of the period). 
Over the test period, the market is more volatile than the previous one, 
and the LSTM-based predictors shows a poor predictive accuracy value of 0.495.
Much of our analysis generated results similar
to those in earlier sections, but interestingly, there is
a positive relation between the magnitude of predictive return 
and the prediction accuracy: that is,
the accuracy consistently increases from 0.529 ($\theta=0.00$) to 0.611 at
($\theta=0.225$), as shown in Table \ref{Threshold_Accuracy2}. 

\begin{table}[htbp]
  \centering
  \caption{Experiment 2 results: Prediction accuracies for the assets whose predictive return is higher than $\theta$ over the test period}
    \begin{tabular}{lrrr}
    \toprule
    $\theta$ & \thead{No. of correct \\ forecasts} & \thead{No. of total\\ forecasts }  & Accuracy \\
    \midrule
    \multicolumn{1}{l}{0} & 142   & 268   & 0.529 \\
    \multicolumn{1}{l}{0.0025} & 124   & 238   & 0.521 \\
    \multicolumn{1}{l}{0.005} & 115   & 218   & 0.527 \\
    \multicolumn{1}{l}{0.0075} & 104   & 192   & 0.541 \\
    \multicolumn{1}{l}{0.01} & 96    & 172   & 0.558 \\
    \multicolumn{1}{l}{0.0125} & 79    & 143   & 0.552 \\
    \multicolumn{1}{l}{0.015} & 73    & 130   & 0.561 \\
    \multicolumn{1}{l}{0.0175} & 63    & 108   & 0.583 \\
    \multicolumn{1}{l}{0.02} & 57    & 95    & 0.6 \\
    \multicolumn{1}{l}{0.0225} & 52    & 85    & 0.611 \\
    \multicolumn{1}{l}{0.025} & 46    & 73    & 0.605 \\
    \bottomrule
    \end{tabular}%
  \label{Threshold_Accuracy2}%
\end{table}%

\section{Applications}\label{sec_app}
Regarding the practical use of TBPs, we provide illustrations 
on how to manage them over multiple-periods and  
and how to incorporate them into an MPT optimization portfolio.

\subsection{TBP Management}

\begin{figure}[t]
\centering
  \scalebox{0.55}
  {
	\includegraphics{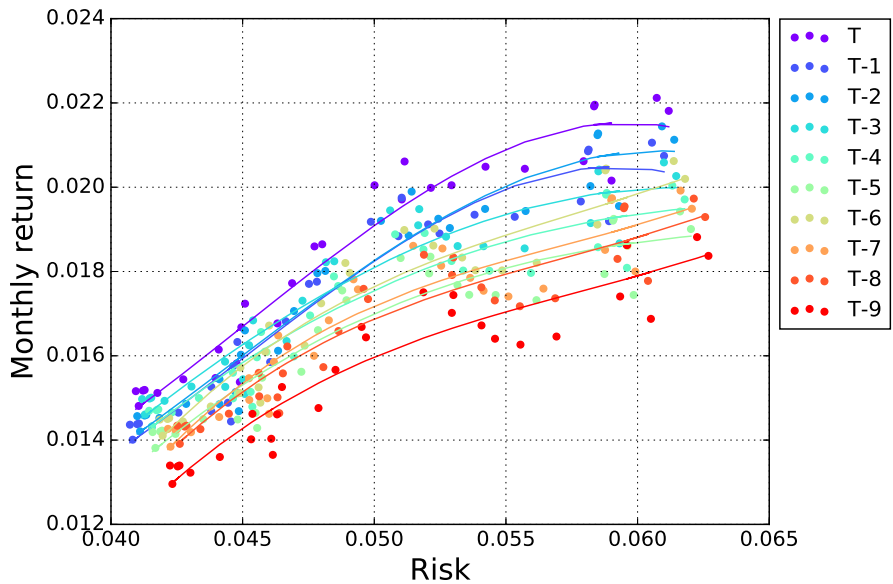}
   }
\caption{(Color online) The realized monthly return versus risk of the TBPs, which are constructed using one-month-ahead return predictions, at $\theta=0.000, 0.0025,\ldots, 0.025$ on the last 10 months of the test period of experiment 1. $T$ is 
the last business month. }
\label{Frontier}
\end{figure}
Figure \ref{Frontier} is a scatter plot of 
the risk--return profiles of the TBPs 
that we are constructed in experiment 1, at different 
$\theta$s ($0.000, 0.0025,\ldots, 0.025$) over the last 10 months, along
with the lines fitted to the polynomial of degree 3.
We will refer to the line as the ``TBP frontiers.''  
Note that the points indicate the realized monthly returns 
and risk of the (predictive) TBPs
built using one-month-ahead predictive returns.
The TBP lines are concave, moving
upward and to the right as $\theta$ increases, thus indicating
that the greater the amount of risk by the increase of $\theta$, the greater 
the realized returns.
This characteristics allows for the design of a TBP with a target risk--return.

To illustrate, let us suppose that an investor at time $T-9$ hopes to build a TBP to achieve
a target risk--return at time $T-8$. 
If the target is the monthly risk of 0.05 
and the monthly return of 0.016, the investor can estimate 
the $\hat{\theta}$, which corresponds to the target from the TBP frontier drawn at $T-9$;
the investor can then build a TBP with the target,
using the predictive return generated and the estimated $\hat{\theta}$ at time $T-9$.
Then, at time $T-8$, the investor will obtain an approximate return of 0.017 and risk of 0.05, 
as seen in the TBP frontier moving upward over the period from time $T-9$ to $T-8$. 
This difference, $0.017-0.015$, is
the estimation risk of the TBP. As seen in the continual shift of TBPs as time passes, 
to maintain a target risk--return, the corresponding $\hat{\theta}$ 
needs to be updated. 
The estimation risk of TBP can be quantified by 
calculating the average of the differences of realization 
and expectation for both return and risk over a past period.
There is the estimation risk, but it seem to be sufficiently small to classify TBPs as 
having different risk aptitude.
The TBP frontier can be more broad, and combined with riskless assets.
In summary, we illustrate the TBP management process:
\begin{enumerate}[itemindent=1cm,label=\bfseries Step \arabic*.]
  \item Set a investment universe (stocks, bonds, ETFs, etc.)
  \item Build forecasting models for future stock return or price
  \item Select a trading position (long-only, short-only, or long--short) 
  and a weighting method (equal-weighted, prediction-weighed, etc.)
 \item (Backtest) Draw the TBP frontiers at different thresholds 
\item Select a target risk--return value and find its corresponding $\hat{\theta}$ on the TBP frontier
\item (Actual investment) Invest in the TBP with the $\hat{\theta}$
\item (Realization) Estimate the TBP at $\hat{\theta}$, and
reinvest in the TBP with the updated $\hat{\theta}$ from the realized TBP frontier
\end{enumerate}

\subsection{Mean-Variance Portfolio}
\begin{figure}[t]
\centering
  \scalebox{0.3}
  {
	\includegraphics{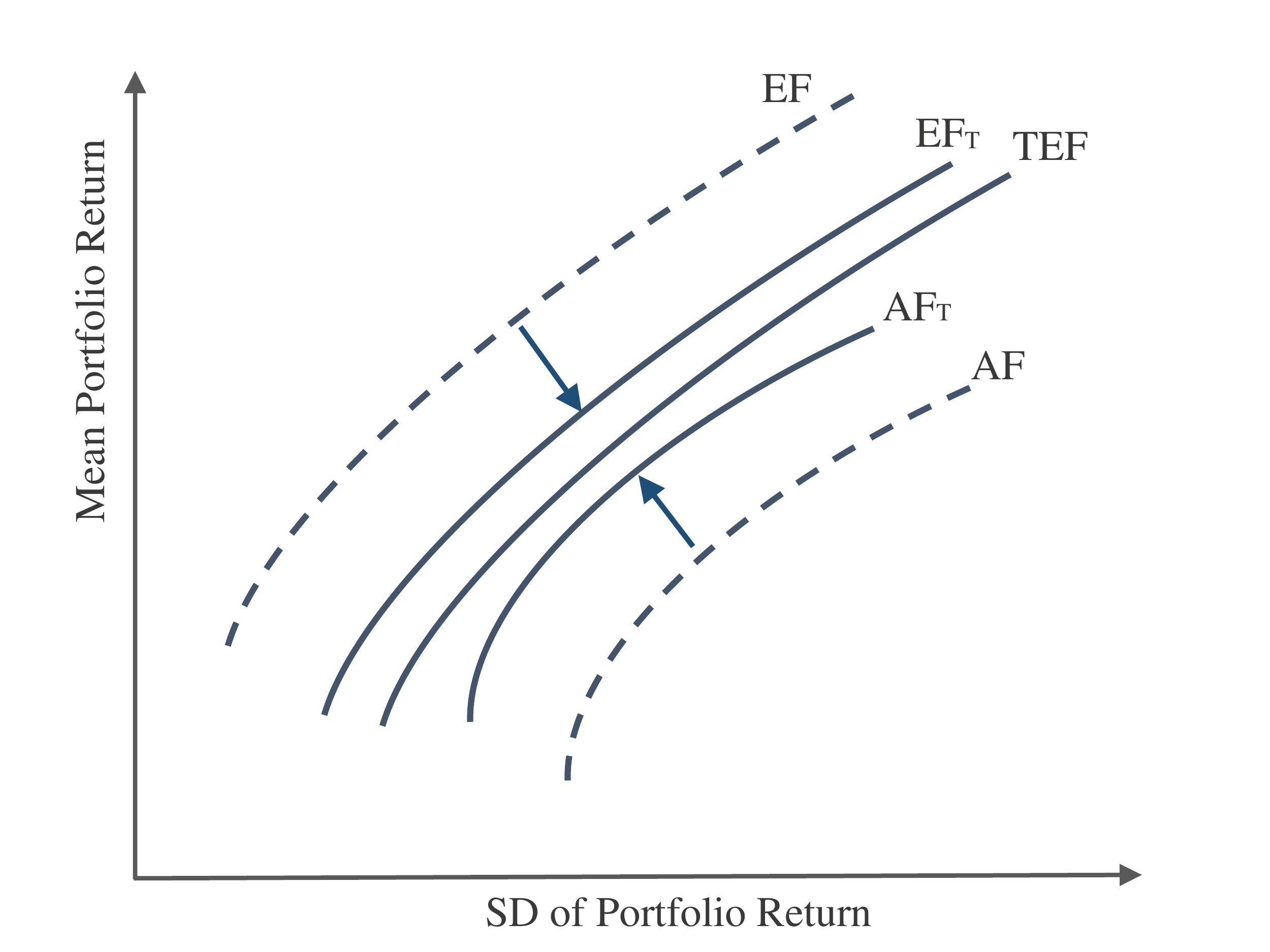}
   }
\caption{(Color online) Schematic scenario for shifting EF to $\textrm{EF}_{T}$ 
by screening the universe at a threshold}
\label{fig_threshold_MPT}
\end{figure}

MPT is a mathematically elegant framework for building a portfolio with 
specific risk-return level.
However, 
it is well known that it is more difficult to estimate the means 
than the covariances of asset returns \cite{mer80},
and errors in the estimates of means will have a greater impact on portfolio weights 
than errors in the estimates of covariances. Furthermore,
as mean--variance optimization is extremely sensitive to expected returns, any
errors therein might make outcomes far from optimal \cite{jor85,bes92}.
For this reason, although theoretical and empirical academic studies 
have examined various MPT aspects, its real-life practical applications 
have mostly focused on minimum variance portfolios.
This estimation error invariably leads to inefficient portfolios,
which can be explained by considering the following three sets of portfolios \cite{bro93}.
\begin{itemize}[label=$\bullet$]
\item True efficient frontier (TEF): the efficient frontier based on true (but unknown
parameters) 
\item Estimated frontier (EF): the frontier based on the estimated (and hence
incorrect) parameters 
\item Actual frontier (AF): the frontier comprising the true portfolio mean and variance points
corresponding to portfolios on the estimated frontier 
\end{itemize}
The use of thresholds on predicted returns
can help mitigate inefficiency by
screening a subset to be predicted more accurately,
as shown in Table \ref{Threshold_Accuracy2}. 
Figure \ref{fig_threshold_MPT} shows a schematic scenario that
$\textrm{EF}_{T}$, which is estimated for the assets 
screened by a threshold,
is located more closely to the TEF. (Here, for simplicity, we ignore the shift in 
EF due to the change in the asset number.)

\section{Conclusion}\label{conclusion}

This study proposes a novel framework by which to construct portfolios 
that target specific risk--return levels. We evaluated 
the RNN networks while examining the hit ratios of the one-month-ahead forecasts 
of stock returns, and then constructed TBPs 
by imposing thresholds for the potential return.
The TBPs are more data-driven in building a portfolio than in existing methods,
in the sense that they are constructed purely on 
the basis of data-driven models by a deep learning technique, in absence of any financial
mathematics or knowledge. 
We showed that the EW of the universe plays the role of the reference portfolio to TBPs, 
and thus serves to quantitatively characterize the TBP.  
The TBP frontiers show that
the threshold is a parameter used to control tradeoff between the risk and the return of portfolios. 
We discussed how to practically manage TBPs 
to maintain a target risk--return over multiple-periods; we also  
discussed the benefit of incorporating
TBPs into MPT.

The TBP is promising, since it provides a simple and straightforward way to
build portfolios with target risk--returns, using predictions alone.
Any prediction model  
can be basically applied to construct TBPs. 
As predictors become more accurate,
TBPs can achieve greater returns, given a certain level of risk. 
In this respect, we believe that
the TBP is a valuable application of machine learning to modern-day investment practice.

\begin{acknowledgements}
This work was partly supported by the ICT R$\&$D program of MSIP/IITP [2017-0-00302, Development of Self Evolutionary AI Investing Technology] and 
the ICT R$\&$D program of MSIP/IITP 
[2014-0-00616, Building an Infrastructure of a Large Size Data Center].
\end{acknowledgements}


\begin{thebibliography}{}


\bibitem{ats09}
G. S. Atsalakis, K. P. Valavanis, Surveying stock market forecasting techniques - Part II: Soft computing methods. Expert Systems with Applications 36(3):5932--5941 (2009)

\bibitem{dix15}
Dixon, M., Klabjan, D.,  Bang, J. H. (2015) Implementing deep neural networks for financial market prediction on the Intel Xeon Phi. 
https://ssrn.com/abstract=2627258 


\newblock
\bibitem{huc09}
Huck, N.
(2009) 
Pairs selection and outranking: An application to the {S}\&{P}100 index.
Journal of Operational Research
196(2):819--825

\newblock
\bibitem{huc10}
Huck, N.
(2010) 
Pairs trading and outranking: The multi-step-ahead forecasting case.
Journal of Operational Research
207(3):1702--1716


\newblock
\bibitem{kra17}
Krauss, C., Do, X. A., Huck, N.
(2017) 
Deep neural networks, gradient-boosted trees, random forests: Statistical arbitrage on the {S}\&{P}500.
259(2):689-702


\newblock
\bibitem{mor14}
Moritz, B., Zimmermann, T.
(2014) 
Deep conditional portfolio sorts: The relation between past and future stock returns.
Working paper

\newblock
\bibitem{ser13}
Sermpinis, G., Theofilatos, K. A., and Karathanasopoulos, A. S., Georgopoulos, E. F.,  Dunis, C. L.
(2013) 
Forecasting foreign exchange rates with adaptive neural networks using radial-basis functions and Particle Swarm Optimization.
European Journal of Operational Research
225(3):528-540


\newblock
\bibitem{tak13}
Takeuchi, L. and Lee, Y.-Y.
(2013) 
Applying deep learning to enhance momentum trading strategies
in stocks.
Working paper

\newblock
\bibitem{cav16}
Cavalcante, R. C., Brasileiro, R. C., Souza, V. L. F., Nóbrega, J. P., Oliveira, A. L. I. (2016) 
Computational Intelligence and Financial Markets: A Survey and Future Directions. 
Expert Systems with Applications 
55:194--211


\newblock
\bibitem{agg17}
Aggarwal, S., Aggarwal, S 
(2017) 
Deep Investment in Financial Markets using Deep Learning Models. 
International Journal of Computer Applications
162(2):40--43

\newblock
\bibitem{gao16}
Gao, T., Li, X., Chai, Y., Tang, Y. 
(2016) 
Deep learning with stock indicators and two-dimensional principal component analysis for closing price prediction system. 
IEEE, Software Engineering and Service Science (ICSESS), 2016 7th IEEE International Conference on

\newblock
\bibitem{zha15}
Zhang, Y.
(2015) 
Using Financial Reports to predict Stock Market
Trends With Machine Learning Techniques.
Oxford University

\newblock
\bibitem{fis17}
Fischer, T., Krauss, C. (2017) Deep learning with long short-term memory networks for financial
market predictions. FAU Discussion Papers in Economics 11/2017, Erlangen. Available at 
http://hdl.handle.net/10419/157808.

\newblock
\bibitem{pan18}
Pang, X., Zhou, Y., Wang, P. et al. J Supercomput (2018). https://doi.org/10.1007/s11227-017-2228-y

\newblock
\bibitem{zin18}
Ziniu Hu, Weiqing Liu, Jiang Bian, Xuanzhe Liu, and Tie-Yan Liu. (2018) Listening to chaotic whispers: A
deep learning framework for news-oriented stock trend prediction. In Proceedings of the Eleventh
ACM International Conference on Web Search and Data Mining, pp. 261–-269. 

\newblock
\bibitem{sin17}
Singh, R. Srivastava, S. (2017) Stock Prediction using Deep Learning, Multimedia Tools and Application,
76 (18), 18569--18584.


\newblock
\bibitem{mar52}
Markowitz, H.
(1952) 
Portfolio Selection.
7(1):77--91


\newblock
\bibitem{fam92}
Fama, E. and French, K. 
(1992) 
The cross-section of expected stock returns. 
Expert Systems with Application 
47:427--465

\newblock
\bibitem{bla92}
Black, F., Litterman, R. 
(1992) 
Global portfolio optimization. 
Financial Analysts Journal 
48(5):28--43

\newblock
\bibitem{mic98}
Michaud, R.
(1998) 
Efficient Asset Management: A Practical Guide to Stock Portfolio
Optimization and Asset Allocation.
Boston: Harvard Business School Press

\newblock
\bibitem{hau91}
Haugen, R., Baker, N.
(1991) 
Pairs selection and outranking: An application to the {S}\&{P}100 index.
Journal of Operational Research
17(3):35--40


\newblock
\bibitem{cho08}
Choueifaty, Y., Coignard, Y.
(2008) 
Toward maximum diversification.
The Journal of Portfolio Management
35(1):40--51


\newblock
\bibitem{qia05}
Qian, E
(2005) 
Risk parity portfolios: Efficient portfolios through true diversification.
Research Paper
https://www.panagora.com/assets/PanAgora-Risk-Parity-Portfolios-Efficient-Portfolios-Through-True-Diversification.pdf

\newblock
\bibitem{qia09}
Qian, E
(2005) 
Risk parity portfolios: The next generation.
Research Paper
https://www.panagora.com/assets/PanAgora-Risk-Parity-The-Next-Generation.pdf



\newblock
\bibitem{fab14}
Faber, M. T. 
(2014) 
A Quantitative Approach to Tactical Asset Allocation. 
The Journal of Wealth Management 
9(4): 69--79

\newblock
\bibitem{kel14}
Keller, W. J., Keuning, J. W.
(2014) 
Momentum, Markowitz, and Smart Beta, A Tactical, Analytical and Practical Look at Modern Portfolio Theory.
https://ssrn.com/abstract=2759734 or http://dx.doi.org/10.2139/ssrn.2759734

\newblock
\bibitem{kel14_2}
Keller, W. J., Bulter, A.
(2014) 
A Century of Generalized Momentum; From Flexible Asset Allocations (FAA) to Elastic Asset Allocation (EAA).
https://ssrn.com/abstract=2543979 or http://dx.doi.org/10.2139/ssrn.2543979

\newblock
\bibitem{kel16}
Keller W. J., Keuning J. W. (2009) Protective Asset Allocation (PAA): A Simple Momentum-Based Alternative for Term Deposits. 
Expert Systems with Applications
36(3):5932--5941


\newblock
\bibitem{cam93}
Campbell, J. Y., Sanford J. G., Jiang, W. (1993)   
Trading volume and serial correlation in stock returns. 
Quarterly Journal of Economics 108: 905--939



\newblock
\bibitem{cho00}
Choueifaty, Y., Coignard, Y. 
(2000) 
Trading volume and cross-autocorrelations in stock returns. 
The Journal of Finance 
55:913--935


\newblock
\bibitem{bao17}
Bao, W., Yue, J., Rao, Y. (2017) A deep
learning framework for financial time series using
stacked autoencoders and long-short term
memory. PLoS ONE 12(7): e0180944. https://doi.
org/10.1371/journal.pone.0180944.

\newblock
\bibitem{fre09}
Freitas, F.D., De Souza, A.F., de Almeida, A.R. (2009), Prediction-based portfolio optimization
model using neural networks, Neurocomputing 72: 2155–-2170.

\newblock
\bibitem{mis16}
Mishra, S., K., Panda, G., Majhi, B. (2016)
Prediction based mean-variance model for constrained portfolio assets
selection using multiobjective evolutionary algorithms.
Swarm and evolutionary computation. 28: 117-–130
%%%%%%%%%%%%%%%%%%%%%%

\newblock
\bibitem{gan13}
Ganeshapillai, G., Guttag, J., Lo, A. (2013) 
Learning connections in financial time series,
in: ICML.



\newblock
\bibitem{elm90}
Elman, Jeffrey L. 
(1990) 
Finding Structure in Time. 
Cognitive Science 
14:179--211


\newblock
\bibitem{ben94}
Bengio, Y. and Simard, P., Frasconi, P. 
(1994) Learning Long-Term Dependencies with Gradient Descent is Difficult. 
IEEE Transactions on Neural Networks 
5(2):157--166


\newblock
\bibitem{hoc97}
Hochreiter, S., Schmidhuber, J. 
(1997) 
Long short-term memory.
Neural Computation 
9(8): 1735--1780

\newblock
\bibitem{cho14}
Cho, K., van Merrienboer B., G{\"{u}}l{\c{c}}ehre {\c{C}}., 
Bougares, F.,  Schwenk, H., Bengio, Y. 
(2014) 
Learning Phrase Representations using {RNN} Encoder-Decoder for Statistical Machine Translation. 
CoRR abs/1406.1078

\newblock
\bibitem{kin14}
Diederik P. K., Jimmy B. 
(2014) 
Adam: {A} Method for Stochastic Optimization. 
http://arxiv.org/abs/1412.6980

\newblock
\bibitem{cho16}
Chollet, F. (2016) Keras: Deep learning library for theano and tensorflow. https://keras.io,
2016.


\newblock
\bibitem{jeg01}
Jegadeesh, N., Titman, S. (2001) Profitability of momentum strategies: an evaluation of
 alternative explanations. Journal of Finance 56: 699--720. 

\newblock
\bibitem{ply01}
Y. Plyakha, R. Uppal, Vilkov, G,. (2012) Why does an equal-weighted portfolio outperform
value-and price-weighted portfolios? Available at SSRN 1787045.

\newblock
\bibitem{mer80}
Merton, R. C. (1980) 
On estimating the expected return on the market: an explatory investigation.
Journal of financial economics 8: 323--361



\newblock
\bibitem{jor85}
Jorion, P.(1985) 
International portfolio diversification with estimation risk.
Journal of business 58(3): 259--278


\newblock
\bibitem{bes92}
Best, M., Grauer, R. (1992) 
Positively weighted minimum-variance portfolios and the structure of asset expected returns.
The journal of financial and quantitative analysis 27(4): 513--537

\newblock
\bibitem{bro93}
Broadie, M., (1993) Computing Efficient Frontiers using Estimated Parameters. 
Annals of Operations Research, 45: 21--–58.





\end{thebibliography}


\end{document}